\documentclass[final,5p,times,twocolumn]{elsarticle}

\usepackage{graphicx}

\usepackage{amssymb}

\journal{Journal of Physics and Chemistry of Solids}

\begin{document}

\begin{frontmatter}



















\title{Interplay of electron-phonon interaction and strong correlations: 
DMFT+$\Sigma$ approach}





\author[1]{M.V. Sadovskii\corref{cor1}}
\ead{sadovski@iep.uran.ru}
\author[1]{E.Z. Kuchinskii}
\author[1]{I.A. Nekrasov}

\address[1]{Institute for Electrophysics, Russian Academy of Sciences, Ural Branch,
Ekaterinburg, Amundsen str. 106, 620016, Russia }


\cortext[cor1]{Corresponding author.}

\begin{abstract}

We discuss interaction of strongly correlated electrons (described within the 
Hubbard model solved by dynamical mean-field theory (DMFT)) with Debye and
Einstein phonons using recently developed DMFT+$\Sigma$ computational scheme. 
Electron-phonon interaction (EPI) is analyzed in adiabatic approximation 
(assuming the validity of Migdal theorem), 
allowing the neglect of EPI vertex corrections. 
This approach is valid for EPI coupling constant $\lambda < \varepsilon_F/\omega_{ph}\sim 10$, 
where $\varepsilon_F$ is Fermi energy and $\omega_{ph}$ is Debye or Einstein frequency.  
For moderate values of  $\lambda$ only small changes in the electronic density 
of states are observed in  DMFT+$\Sigma$ approximation for both weakly and 
strongly correlated metallic regimes. Metal-insulator (Mott) transition due to 
the increase of  Hubbard interaction U is slightly inhibited by EPI. 
Our main aim is to discuss the interplay of  ``kinks'' in electronic 
dispersion due to EPI and recently discovered kinks of electronic origin. 
For the certain region of model parameters coexistence of 
phonon ``kinks'' in  electronic dispersion with purely electronic ``kinks'' is 
readily observed and we formulate some simple criteria of such coexistence. 
However, for most general combinations of model parameters phonon ``kinks'' 
make electronic ``kinks'' hardly observable.
In the general case an increase of 
Hubbard interaction U rapidly suppresses the slope of electronic dispersion 
within the phonon ``kink''.  These results are important for deeper 
understanding of the  shape and  evolution of electronic dispersions in  
strongly correlated systems such as copper oxides, where different kinds of 
``kinks'' were recently observed in ARPES experiments.

\end{abstract}


\end{frontmatter}



\section{Introduction}

The problem of the interplay of strong electronic correlations with electron--phonon
interaction is of central importance in the physics of highly
correlated systems. Actually there is rather long history of such studies, e.g.
one of the most popular models for electron-phonon interaction (EPI)
in strongly correlated systems is the so-called Hubbard-Holstein model (HHM).
The Hubbard model \cite{Hubbard} itself describes local Coulomb interaction of 
electrons on a lattice including e.g. Mott-Hubbard metal-insulator transition.
On the other hand Holstein model contains local linear displacement-to-density
interaction of conducting electrons with local (Einstein) phonon modes
\cite{Holstein59}.

Active investigations of the properties of the HHM were undertaken in the framework
of dynamical mean-field theory (DMFT) \cite{DMFT_method}, which is non-perturbative approach
with respect to interaction parameters of the Hubbard model.
Among many others one should mention DMFT solution of HHM for the case where impurity
solver used was the numerical renormalization group (NRG) \cite{NRGrev}.
The mapping of HHM to Anderson-Holstein impurity was first
performed by Hewson and Mayer \cite{HewMay02}. It was shown that using NRG one can
compute in a numerically exact manner total electron-phonon contribution
to the self-energy of the problem, thus making solution of the HHM
non-perturbative also with respect to electron-phonon coupling strength.

However, up to now there are apparently no studies of strongly correlated electrons
interacting with Debye phonons. It is even more surprising in view of the widely
discussed physics of kinks in electronic dispersion observed in ARPES experiments
40-70 meV below the Fermi level of high-temperature superconductors \cite{Lanz}, which are
often attributed to EPI \cite{Shen}. To our knowledge problem of kink formation on electronic 
dispersion caused by EPI in strongly correlated systems was briefly discussed 
within HHM in papers by Hague \cite{Hague} and Koller~{\it et al.} \cite{Koller05}.

In this paper 
we consider the influence of Debye or Einstein phonons on the weakly and
strongly correlated electrons within our recently developed DMFT+$\Sigma$ 
approach, studying electron dispersion and density of
states (DOS), in particular close to Mott-Hubbard metal insulator transition.
We analyze in details how EPI affects electronic dispersions in correlated
metal and discuss the interplay of recently discovered kinks of purely
electronic nature in electronic dispersion \cite{Nature} and usual phonon 
kinks in the electronic spectra.

\section{
DMFT+$\Sigma$ computational details}
\label{comp}

The major assumption of our DMFT+$\Sigma$ approach is that the lattice
and time Fourier transform of the single-particle Green function 
can be written as:
\begin{equation}
G_{\bf p}(\varepsilon)=\frac{1}{\varepsilon+\mu-\varepsilon({\bf p})-\Sigma(\varepsilon)
-\Sigma_{\bf p}(\varepsilon)}
\label{Gk}
\end{equation}
where $\varepsilon({\bf p})$ is the bare electron dispersion,
$\Sigma(\varepsilon)$ is the {\em local} self--energy of DMFT, 
while $\Sigma_{\bf p}(\varepsilon)$ is some 
``external'' (in general case momentum dependent) self--energy. 
Advantage of our generalized approach is the additive form of 
the self-energy (neglect of interference) in Eq. (\ref{Gk}) 
\cite{fsdistr,dmftsk,opt}. 
It allows one to keep the set of self-consistent equations of standart DMFT 
\cite{DMFT_method}. However there are two distinctions. 
First, on each DMFT iterations we recalculate corresponding  ``external'' 
self-energy $\Sigma_{\bf p}(\mu,\varepsilon,[\Sigma(\varepsilon)])$ within some 
(approximate) scheme, taking into account interactions e.g. with collective 
modes (phonons, magnons etc.) or some order parameter fluctuations.
Second, the local Green's function of effective impurity problem is defined as
\begin{equation}
G_{ii}(\varepsilon)=\frac{1}{N}\sum_{\bf p}\frac{1}{\varepsilon+\mu
-\varepsilon({\bf p})-\Sigma(\varepsilon)-\Sigma_{\bf p}(\varepsilon)},
\label{Gloc}
\end{equation}
at each step of the standard DMFT procedure.

Eventually, we get the desired Green function in the form of (\ref{Gk}),
where $\Sigma(\varepsilon)$ and $\Sigma_{\bf p}(\varepsilon)$ are those 
appearing at the end of our iteration procedure.

To treat electron-phonon interaction for strongly correlated system we
just introduce $\Sigma_{\bf p}(\varepsilon)=\Sigma_{ph}(\varepsilon,{\bf p})$ due
to electron--phonon interaction within the usual Fr\"ohlich model.
To solve single impurity Anderson problem we use NRG\cite{NRGrev}.
All calculations are done at nearly zero temperature and at half filling.
For ``bare'' electrons we assume semielliptic DOS with half--bandwidth $D$.

According to the Migdal theorem in adiabatic approximation \cite{Migdal} we can 
restrict ourselves with the simplest first order contribution to 
$\Sigma_{ph}(\varepsilon,{\bf p})$, neglecting vertex 
corrections due electron-phonon coupling which are small over adiabatic 
parameter ${(\omega_{D},\omega_{0}})/{\varepsilon_F}\ll 1$ \cite{Migdal}:
\begin{eqnarray}
\Sigma_{ph}(\varepsilon,\textbf{p})=ig^2 \sum_{\omega,\textbf{k}}
\frac{\omega^2_0(\textbf{k})}{\omega^2-\omega^2_0(\textbf{k})+i\delta}\times
\nonumber\\
\frac{1}{\varepsilon+\omega+\mu-\varepsilon({\textbf{p}+\textbf{k}})-
\Sigma(\varepsilon+\omega)-\Sigma_{ph}(\varepsilon+\omega,\textbf{p}+\textbf{k})}
\label{phse}
\end{eqnarray}
where $g$ is the usual electron-phonon interaction constant, 
$\omega_0(\textbf{k})$ is phonon dispersion, which in our case is taken 
as in the standard Debye or Einstein model
\begin{equation}
\omega_0(\textbf{k})=\left\{ \begin{array}{ccc}uk,&\quad &k < \frac{\omega_D}{u}\\
\omega_0,&\quad &k <k_0 \end{array} \right. .
\label{dspec}
\end{equation}
Here $u$ is the sound velocity, $\omega_D$, $\omega_0$ are Debye and Einstein 
frequencies with cut-off $k_0$ of the order of Fermi momentum $p_F$.

Actually $\Sigma_{ph}(\varepsilon,\textbf{p})$ defined by Eq.~(\ref{phse})
has weak momentum dependence which we can omit and continue only with
significant frequency dependence. For Debye spectrum of phonons 
Eq.~(\ref{phse}) can be rewritten as (cf. similar analysis in Ref. \cite{diagrammatics})
\begin{eqnarray}
&&\Sigma_{ph}(\varepsilon)=\frac{-ig^2}{4\omega_c^2} \int_{-\infty}^{+\infty}\frac{d\omega}{2\pi}
\bigl\{\omega_D^2+\omega^2ln\bigl|\frac{\omega_D^2-\omega^2}{\omega^2}\bigr|+\nonumber\\
&&+i\pi\omega^2\theta(\omega_D^2-\omega^2)\bigr\}I(\varepsilon+\omega)
\label{phdb}
\end{eqnarray}\
with a characteristic frequency $\omega_c\!=\!p_Fu$ of the order of $\omega_D$,
while for Einstein spectrum:
\begin{eqnarray}
\Sigma_{ph}(\varepsilon )=\frac{ig^2 k^2_0}{16\pi p^2_F} 
\Bigl\{&&\!\!\!\!\!\!\!\!\!\!\!\!\! -i\pi (I(\varepsilon\!+\!\omega_0)\!+\!I(\varepsilon\! -\!\omega_0))+ \\
\!\!\!\!\!\!\!\!\!\!\!\int\limits_{0}^{\infty}\!\!\! \frac{d\omega}{\omega}
(I(\varepsilon+\omega_0\!\!+\!\omega )\!+\!
 I(\varepsilon-\!\!\!\!\!\!\!\!\!\!\!&&\!\omega_0\!\!-\!\omega )
\!-\!I(\varepsilon\!+\!\omega_0\!\!-\!\omega )\!-\!I(\varepsilon\!-\!\omega_0\!\!+\!\omega ))
\Bigr\}\nonumber
\label{phin}
\end{eqnarray}
with
\begin{equation}
I(\epsilon)=\int_{-D}^{+D}d\xi\frac{N_0(\xi)}{E_{\varepsilon}-\xi}.
\end{equation}
where $E_{\varepsilon}=\varepsilon-\Sigma(\varepsilon)-\Sigma_{ph}(\varepsilon)$.
For the case of semielliptic non-interacting DOS $N_0(\varepsilon)$ with
half-bandwidth $D$ we get:
\begin{equation}
I(\epsilon)=\frac{2}{D^2}(E_{\varepsilon}-\sqrt{E_{\varepsilon}^2-D^2}),
\end{equation}
It is convenient to introduce the dimensionless electron-phonon coupling constant 
as\cite{diagrammatics}:
\begin{equation}
\lambda_D=g^2N_0(\varepsilon_F)\frac{\omega_D^2}{4\omega_c^2},~~
\lambda_E=g^2N_0(\varepsilon_F)\frac{k_0^2}{4 p_F^2}.
\label{lambda}
\end{equation}
To simplify our analysis we shall not perform fully self-consistent calculations 
neglecting phonon renormalization due to EPI\cite{diagrammatics}, assuming that 
the phonon spectrum (\ref{dspec}) is fixed by the experiment.

\section{Results and discussion}
\label{results}

\begin{figure}[t]
\begin{center}
\includegraphics[scale=0.3]{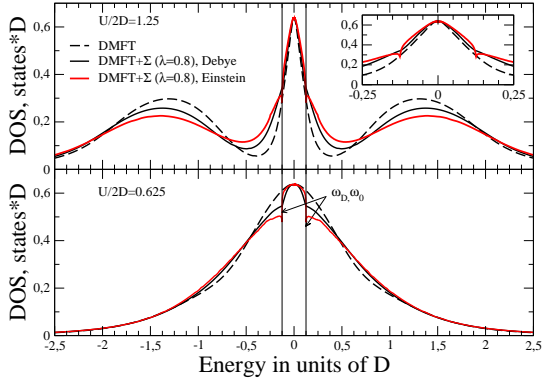}
\end{center}
\vspace{-0.7cm}
\caption{ 
Comparison of DOSes obtained within standard DMFT (dashed lines) and 
DMFT+$\Sigma_{ph}$ methods (solid lines) for Debye (black) and Einstein (red) 
phonons for strong (upper panel, U/2D=1.25) 
and weak (lower panel, $U/2D=$0.625) Hubbard interaction regimes.
Dimensionless electron--phonon coupling constant $\lambda_{D}=\lambda_{E}$=0.8.
\label{DOSes}
}
\end{figure}

Let us start from comparison between pure DMFT and DMFT+$\Sigma_{ph}$ 
DOSes for strong (U/2D=1.25) and weak (U/2D=0.625) Hubbard interaction
presented in Fig.~\ref{DOSes} on upper and low panels correspondingly.
Dimensionless EPI constant (\ref{lambda}) used in these calculations 
was $\lambda_{D}=\lambda_{E}$=0.8, while Debye and Einstein frequencies  were taken to be
$\omega_D$=$\omega_0$=0.125D. 
In both cases we observe some spectral weight redistribution due to EPI.
For U/2D=1.25 (upper panel of Fig.~\ref{DOSes})
we see the well developed three peak structure typical for strongly correlated 
metals. In the energy interval {\bf{$\pm \omega_D,\omega_0$}} around the Fermi energy
(which is taken as zero energy at all figures below) 
there is almost no difference in the DOS quasiparticle peak line shape obtained 
from pure DMFT and DMFT+$\Sigma_{ph}$.
However outside this interval DMFT+$\Sigma_{ph}$ quasiparticle peak becomes 
significantly broader with spectral weight coming from Hubbard bands
and it is  more pronounced for the case of Einstein phonons.
This broadening of DMFT+$\Sigma_{ph}$ quasiparticle peak leads
as we show below to inhibiting of metal to insulator transition.
In the case of U/2D=0.625 there are no clear Hubbard bands formed but only some 
``side wings'' are observed. 
Spectral weight redistribution on the lower panel of Fig.~\ref{DOSes} is not 
dramatic, though qualitatively different from the case of U/2D=1.25. 
Namely, main deviations between pure DMFT and DMFT+$\Sigma_{ph}$ happen in the 
interval $\pm \omega_D$, where one can observe kind of ``cap'' in 
DMFT+$\Sigma_{ph}$  DOS. Corresponding spectral weight goes to
the energies around $\pm$U, where Hubbard bands are supposed to form.
The lineshape of the ``cap'' is slightly different for the Debye and Einstein
phonons due to different behavior of $\Sigma_{ph}$ (Eqs.~(\ref{phdb}),
(\ref{phin})) at energies $\pm \omega_D,\omega_0$. For Einstein phonons  
Im$\Sigma_{ph}$ at these energies sharply drops down to zero. This leads to 
sharp cusps of DOS at $\pm \omega_0$ as shown at the insert in Fig.~\ref{DOSes}.

\begin{figure}[t]
\begin{center}
\includegraphics[scale=0.3]{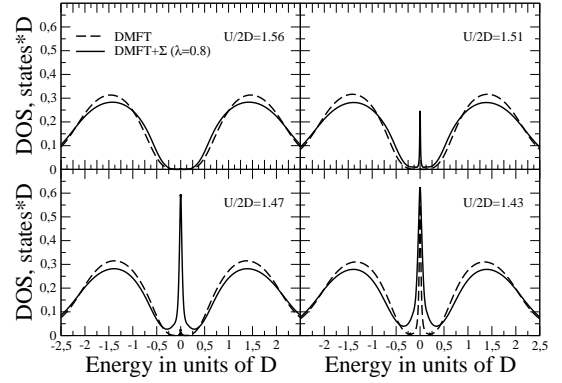}
\end{center}
\vspace{-0.7cm}
\caption{ 
Sequence of DOSes obtained within standard DMFT (dashed lines) and 
DMFT+$\Sigma_{ph}$ for Debye model
(solid lines) methods close to metal-insulator transition 
(from top-left to bottom right) with $\lambda_{D}$=0.8.
\label{MITDOSes}}
\end{figure}

In Fig.~\ref{MITDOSes} we compare the behavior of pure DMFT and 
DMFT+$\Sigma_{ph}$ DOSes for different U/2D values close to Mott-Hubbard 
metal-insulator transition for the case of Debye phonon spectrum.
For U/2D=1.56 both standard DMFT and DMFT+$\Sigma_{ph}$ produce insulating 
solution. However there is some difference between these solutions. 
The DMFT+$\Sigma_{ph}$ Hubbard bands are lower and broader than DMFT ones
because of additional interaction (EPI) included.
With decrease of U for U/2D=1.51 and 1.47 we observe that DMFT+$\Sigma_{ph}$ results correspond to metallic 
state (with narrow quasiparticle peak at the Fermi level), 
while conventional DMFT still produces insulating solution.
Only around U/2D=1.43 both DMFT and DMFT+$\Sigma_{ph}$ results turn out to be 
metallic. Overall DOSes lineshape is the same as discussed above.
These results show that with the increase of U finite EPI slightly inhibits 
Mott-Hubbard transition from metallic to insulating
phase. For the case of Einstein phonons the MIT is inhibited even stronger.
This result is similar to what was observed for the HHM in weak EPI 
regime\cite{KolMayHew04,Jeon,Mayer04}.
For more deep insight into these results on DOS we have also analyzed
the fine structure of corresponding self-energies $\Sigma(\varepsilon)$
and $\Sigma_{ph}(\varepsilon)$. Relevant details can be found in Ref. \cite{prb}.

Now we address the issue of a sudden change of the slope of electronic 
dispersion, the so-called kinks.
It is well known that interaction of electrons with some bosonic mode
always produces such a kink. In the case of EPI typical kink energy is 
just the Debye  $\omega_D$ or Einstein $\omega_0$ frequency.
Kinks of purely electronic nature were recently reported in 
Ref.~\cite{Nature}. 
The energy of purely electronic kink as derived in Ref.~\cite{Nature} 
for semielliptic bare DOS is given by
\begin{equation}
\omega^*=Z_{FL}(\sqrt{2}-1)D,
\label{wst}
\end{equation}
where D is the half of the bare bandwidth of electrons and 
$Z_{FL}=(1-\frac{\partial Re\Sigma)}{\partial \varepsilon}\bigr|_{\varepsilon=
\varepsilon_F})^{-1}$ is Fermi liquid quasiparticle weight.
The rough estimate of $\omega^*$ is given by the half-width of quasiparticle 
peak of DOS at its half-height.

Our calculations clearly demonstrate that electronic kinks are hardly observable
on the background of phonon kinks and special care should be taken to separate
them by rather fine tuning of the parameters of our model. To clarify this
situation we introduce an additional characteristic of the kink --- the shift of
electron dispersion in momentum space $\delta p$ at kink energy.
From simple geometry we estimate for phonon kinks
\begin{equation}
\delta p_{ph}=\frac{(\omega_D,\omega_0)}{v_F}\lambda_{D,E}
\label{dpph}
\end{equation}
where $v_F$ is the bare Fermi velocity and $\lambda_{D,E}$ was defined in 
Eq.~(\ref{lambda}).
For electronic kink the similar estimate is
\begin{equation}
\delta p_{e}=\frac{\omega^*}{v_F^*}\bigl(1-\frac{Z_{FL}}{Z_{0}}\bigr)\equiv
\frac{\omega^*}{v_F^{*}}\lambda_e,
\label{dpe}
\end{equation}
where $Z_0$ is quasiparticle weight in the case of absence of electronic kinks
(the same as $Z_{cp}$ defined in Ref.~\cite{Nature}). 
Velocity $v_F^*$ is the Fermi velocity of initial dispersion, 
but it can not be just a bare one.
As was reported in Ref.~\cite{Nature} 
electronic kinks can be observed only for
rather strong Hubbard interaction when three peak structure 
in the DOS is well developed and electronic
dispersion is strongly renormalized by correlation effects.
This renormalization is determined by $\lambda_e$ defined in Eq.~(\ref{dpph}),
which can be seen as kind of dimensionless interaction constant. 
In the case when both slopes on the Fermi level and out of $\pm \omega^*$ 
energy interval are equal there will be no electronic kink at all.

\begin{figure}[t]
\begin{center}
\includegraphics[scale=0.35]{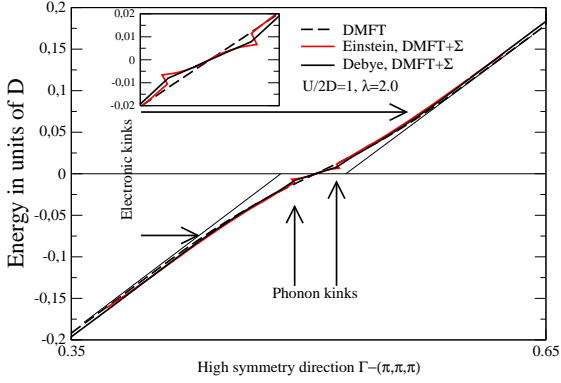}
\end{center}
\vspace{-0.7cm}
\caption{(Color online)
Quasiparticle dispersions obtained from standard DMFT 
(black dashed line) and DMFT+$\Sigma_{ph}$ for Debye (solid black line)
and Einstein (solid red line) phonons 
along the part of high symmetry direction $\Gamma-(\pi,\pi,\pi)$. 
\label{ephkinks}}
\end{figure}

Now  we can choose parameters of our model
to make both kinks simultaneously visible.
First of all one should take care that $\omega_D \ll \omega^*$.
For U/2D=1 with U=3.5~eV we get $\omega^*\sim$ 0.1D and a reasonable
value of Debye (or Einstein) frequency is $\omega_D( \mbox{or}\ \omega_0)\sim$ 0.01D.
To make phonon kink pronounced at such relatively low Debye (or Einstein) frequency
(cf. Eq.~(\ref{dpph})) we have to increase EPI constant and
we take $\lambda_{D}=\lambda_{E}$=2.0.
To demonstrate coexistence of both these types of kinks we plot
the energy dispersion of simple cubic lattice with nearest neighbors 
transfers only, along the high symmetry direction $\Gamma-(\pi,\pi,\pi)$ \cite{Nature}.
In Fig.~\ref{ephkinks} we show dispersion along this direction close to the 
Fermi level. 
The difference of lineshapes of Debye and Einstein kinks is illustrated
at the insert in the Fig.~\ref{ephkinks}. As discussed above ``Einstein'' kink 
is more sharp.

\begin{figure}[ht]
\begin{center}
\includegraphics[scale=0.35]{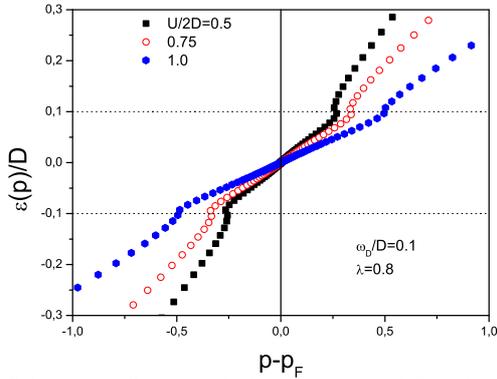}
\end{center}
\vspace{-1.1cm}
\caption{(Color online)
Quasiparticle dispersions around Fermi level
with Debye phonon kinks obtained from DMFT+$\Sigma_{ph}$ calculations
for different interaction strengths U/2D=\ 0.5,\ 0.75,\ 1.0;
\ $\lambda_D=0.8$, $\omega_D$=0.1D. 
\label{Ukinks}}
\end{figure}

Finally we address to the behavior of phonon kinks in electronic spectrum
as function of Hubbard interaction U. As U/2D ratio grows Fermi velocity in 
Eq.~(\ref{dpph}) goes down, so that momentum shift of kink position $\delta p$ 
moves away from $p_F$, while kink energy remains at $\omega_D$. This is
confirmed by our direct DMFT+$\Sigma_{ph}$ calculations producing the
overall picture of spectrum evolution shown in Fig.~\ref{Ukinks}. 

\section{Conclusion}
\label{conclusion}

This work is a first attempt to analyze strongly correlated electrons, treated
within DMFT approach to the Hubbard model, interacting with either Debye or 
Einstein phonons. EPI was treated within the simplest (Migdal theorem) approach 
in adiabatic approximation, allowing the neglect of vertex corrections. 
DMFT+$\Sigma_{ph}$ approach allows us to use the standard momentum space 
representation for phonon self-energy (\ref{phse}), while the general structure 
of DMFT equations remains intact.

Mild EPI leads to rather insignificant changes of electron density of states,
both in correlated metal and in Mott--insulator state, slightly inhibiting
metal to insulator transition with increase of U.
However, kinks in the electronic dispersion due to EPI dominate for the most
typical values of the model parameters, making kinks of purely electronic
nature, predicted in Ref. \cite{Nature}, hardly observable. Special care (fine
tuning) of model parameters is needed to separate these anomalies in
electronic dispersion in strongly correlated systems.
We have also studied phonon kinks evolution with the strength of electronic
correlations demonstrating the significant drop in the slope of electronic 
dispersion close to the Fermi level with the growth of Hubbard interaction $U$.
Quantitative difference of results for the cases of Debye and Einstein phonon 
spectra was observed both in DOS and kink behavior in electronic dispersion.




This work is partly supported by RFBR grant 08-02-00021 and was performed
within the framework of programs of fundamental research of the Russian Academy
of Sciences (RAS) ``Quantum physics of condensed matter'' (09-$\Pi$-2-1009) and 
of the Physics Division of RAS  ``Strongly correlated electrons in solid states'' 
(09-T-2-1011). IN thanks Grant of President of Russia MK-614.2009.2,
interdisciplinary UB-SB RAS project, and Russian Science Support
Foundation.









\end {document}